\documentstyle[sprocl]{article}

\def\be{\begin{equation}}
\def\ee{\end{equation}}
\def\bea{\begin{eqnarray}}
\def\eea{\end{eqnarray}}

\begin{document}
\title{NON-SPECTATOR CONTRIBUTIONS TO $B\to X_s\eta'$ DECAYS}
\author{ M.R. AHMADY }
\address{LINAC Laboratory, The Institute of Physical and Chemical Research (RIKEN),\\
2-1 Hirosawa, Wako, Saitama 351-01, Japan}
\author{ E. KOU, A. SUGAMOTO}
\address{Department
of Physics, Ochanomizu University,\\
1-1 Otsuka 2, Bunkyo-ku, Tokyo 112, Japan}
%
%

\maketitle
\abstracts{We propose a nonspectator mechanism in which $\eta'$ is produced by two gluon fusion as the underlying process to explain the inclusive $B\to X_s\eta' $ as well as the  exclusive $B\to K^{(*)}\eta' $ decays.}

The CLEO collaboration has recently discovered an unexpectedly large branching ratio for the semi-inclusive hadronic $B\to X_s\eta'$ decay\cite{cleo}:
\begin{equation}
BR(B\to X_s\eta' )=(7.5\pm 1.5\pm 1.1)\times 10^{-4}\;\;\;2.0\le p_{\eta'}\le 2.7\;\; GeV\;\;.
\end{equation}
The corresponding exclusive decay rate has also been measured:
\begin{equation}
BR(B\to K\eta' )=(7.8 ^{+2.7}_{-2.2}\pm 1.0)\times 10^{-5}\;\; .
\end{equation}
Possible mechanisms behind this large production of fast $\eta'$ meson have been discussed in recent papers\cite{as,hz,yc,ht,kp}.  In this paper, we investigate the possibility that a somewhat different process might be the underlying mechanism for $B\to X_s\eta'$.  We propose a non-spectator process in which $\eta'$ is produced via fusion of the gluon from QCD penguin $b\to sg^*$ and another one emitted by the light quark inside B meson.  We also calculate the branching ratios $BR(B\to K\eta' )$ and $BR(B\to K^*\eta' )$ in the context of factorization.  The effective Hamiltonian can be written as:
\begin{equation}
H_{eff}=CH(\bar s\gamma_\mu (1-\gamma_5)T^ab)(\bar q\gamma_\sigma T^a q)\frac{1}{p^2-M_g^2}\epsilon^{\mu\sigma\alpha\beta}q_\alpha p_\beta\;\; ,
\end{equation}
where
\begin{equation}
C=\lambda_t\frac{G_F}{\sqrt{2}}\frac{\alpha_s}{2\pi}E_0\;\; ,
\end{equation}
and the effective gluon mass $M_g$ is due to bound state effects.  Alternatively, one may use the usual gluon propagator (no effective mass) along with a model which incorporates the binding effects and off-shellness of the light quark inside B meson.  H is the form factor parametrizing $gg\eta'$ vertex. In writing (3) we considered only the dominant chromo-electric operator.  A re-arrangement of (3) via Fierz transformation and using the definition for the $B$ meson decay constant $f_B$ results in:
\begin{equation}
<\eta' X_s|H_{eff}|B>=\frac{Cf_BH}{9(p^2-M_g^2)}\left [-(\bar s\gamma_\sigma\gamma_\rho\gamma_\mu (1-\gamma_5)q)p_B^\rho+\left (\frac{M_B^2}{m_q+m_b}\right )(\bar s\gamma_\sigma\gamma_\mu (1+\gamma_5)q)
\right ]\epsilon^{\mu\sigma\alpha\beta} q_\alpha p_\beta\;\; .
\end{equation}
Hereafter, we take the light quark mass $m_q=0$.
Using the usual convention for the invariant variables $s={(p_{\eta'}+k' )}^2$, $t={(p_s+k' )}^2$ and $u={(p_s+p_{\eta'})}^2$.
\begin{eqnarray}
\nonumber\frac{d\Gamma (B\to X_s\eta' )}{dtdu}&=&\displaystyle \frac{C^2f_B^2H^2}{648\pi^3M_B^3{(p^2-M_g^2)}^2}\times \\
\nonumber\displaystyle
&&\left [p^2X\left \{ (W-Y-\frac{p^2}{2})(W-X)-XZ+(X-\frac{s+p^2}{2})W\right \} \right . \\
\nonumber\displaystyle &-&q^2ZW^2+XYZW+(s-2Y-q^2)(X-\frac{s+p^2}{2})W^2  \\
\nonumber\displaystyle &-&{\left (\frac{M_B^2}{m_b}\right )}^2\left \{ p^2(X-\frac{s+p^2}{2})(s-Y-q^2)-p^2q^2(W+Z-Y-\frac{p^2}{2}) \right . \\
\displaystyle &+& \left . \left . Y^2Z-(s-2Y-q^2)(W-Y-\frac{p^2}{2})Y\right \}
\right ]\;\; ,
\end{eqnarray}
where $W=(u-m_b^2)/2$, $X=(m_b^2-m_s^2+s)/2$, $Y=(m_{\eta'}^2-p^2-q^2)/2$ and $Z=(t-m_s^2)/2$.  $m_{X_s}^2=t$ is the invariant mass of the final state strange hadron.  The differential decay rate (6) depends on the virtualities of the internal gluons both explicitly and implicitly through the form factor $H$.  However, $H(p^2,q^2,m_{\eta'}^2)$ is suppressed for large values of $p^2$ and $q^2$.  Therefore, the dominant contribution to the decay rate is expected to arise from small virtuality region.  On the other hand, in the non-spectator mechanism, due to kinematical freedom, one can impose a constraint such as $p^2=0$.  Consequently, $q^2$ can be expressed as:
\begin{equation}
q^2=m_b^2+m_{\eta'}^2-u+(t-m_s^2)\frac{u-m_b^2}{m_B^2-u}\;\; .
\end{equation}

For our numerical evaluations, we have taken $m_b=4.5$ GeV, $m_s=0.15$\footnote{For phase space calculation, $m_s$ is taken to be the constituent quark mass $m_s^{\rm constituent}\approx 0.45$ GeV.}, $M_g\approx \Lambda_{QCD}\approx 0.3$ GeV (see our explanation following eqn. (3)), $\alpha_s=0.2$, $f_B=0.2$ GeV and $\vert V_t\vert=\vert V_{tb}V_{ts}^*\vert\approx\vert V_{cb}\vert\approx 0.04$ .  In order to obtain the total branching ratio with the experimental cut \mbox{$2.0\leq p_{\eta'}\leq 2.7$ GeV}, the differential decay rate is integrated over the range \mbox{$m_s^{\rm constituent}=0.45\leq m_{X_s}\leq 2.32$ GeV} resulting in:
\begin{equation}
BR(B\to X_s\eta' )=4.7\times 10^{-3}\;\;\; 2.0\leq p_{\eta'}\leq 2.7\;\;{\rm GeV}\;\; .
\end{equation}
The momentum dependence of $H(q^2,0,m_{\eta'}^2)$ has not been taken into account in the above estimate.  However, even if one considers up to an order of magnitude suppression due to this form factor, our result indicates that the non-spectator mechanism is indeed the dominant process making up the bulk of the experimental data (1).
Using eqn. (5) in conjunction with the factorization assumption and the definition of the decay constants $f_K$ and $f_{K^*}$ for $K$ and $K^*$ respectively, we obtain 
\begin{equation}
<\eta' K|H_{eff}|B>=-i\frac{CHf_Bf_K}{9(p^2-M_g^2)}\left (p_B.qp_K.p-p_B.pp_K.q\right )\;\; ,
\end{equation}
leading to the exclusive decay rate
\begin{equation}
\Gamma (B\to K\eta' )=\frac{C^2H^2f_B^2f_K^2}{1944\pi M_g^4}{\vert \vec p_K\vert}^3\left (m_{\eta'}^2+4{\vert \vec p_K\vert}^2\right )p_0^2\;\; ,
\end{equation}
and 
\begin{eqnarray}
<\eta' K^*|H_{eff}|B>=\displaystyle -i\frac{CHf_Bf_{K^*}}{9(p^2-M_g^2)}\left [p_B.q\epsilon .p-p_B.p\epsilon .q \right . \\
\nonumber\displaystyle\left . +\frac{2m_B^2m_s}{m_bm_{K^*}^2}\left (-i\epsilon^{\mu\sigma\alpha\beta}{p_{K^*}}_\mu\epsilon_\sigma q_\alpha p_\beta +p_{K^*}.q\epsilon .p-p_{K^*}.p\epsilon .q\right )
\right ]\;\; ,
\end{eqnarray}
resulting in
\begin{eqnarray}
\nonumber\Gamma (B\to K^*\eta' )&=&\frac{C^2H^2f_B^2f_{K^*}^2}{1296\pi M_g^4m_B^2}\vert\vec p_{K^*}\vert \int^1_{-1}F(x)dx\;\;, \\
\nonumber
F(x)&=&{(C_1+C_3C_5)}^2\frac{C_4^2}{m^2_{K^*}}+{(C_2+C_4C_5)}^2(-q^2+\frac{C_3^2}{m_{K^*}^2}) \\
&-&2(C_1+C_3C_5)(C_2+C_4C_5)(-C_6+\frac{C_4C_3}{m_{K^*}^2}) \\
\nonumber
&+&C_5^2\left [m_{K^*}^2C_6^2-C_3C_6C_4-C_4(C_3C_6-q^2C_4)\right ]\;\; '
\end{eqnarray}
where
\begin{eqnarray}
\nonumber C_1&=&\frac{m_B^2+m_{\eta'}^2-m_{K^*}^2}{2}-C_2\;\; ,\\
\nonumber C_2&=&m_Bp_0\;\; , \\
\nonumber C_3&=&\frac{m_B^2-m_{\eta'}^2-m_{K^*}^2}{2}-C_4\;\; , \\
\nonumber C_4&=&\left [{(m_{K^*}^2+{\vert\vec p_{K^*}\vert}^2)}^{1/2}-\vert\vec p_{K^*}\vert x\right ]p_0\;\; , \\
\nonumber C_5&=&\frac{2m_sm_B^2}{m_bm_{K^*}^2}\;\; , \\
\nonumber C_6&=&\frac{m_{\eta'}^2-q^2}{2}\;\; , \\
\nonumber \vert\vec p_{K^*}\vert &=&\frac{{[(m_B^2-{(m_{\eta'}+m_{K^*})}^2)(m_B^2-{(m_{\eta'}-m_{K^*})}^2)]}^{1/2}}{2m_B}\;\; .
\end{eqnarray}

Eqns. (10) and (12) are derived by imposing the $p^2=0$ constraint.  $\vert\vec p_K\vert$ ($\vert\vec p_{K^*}\vert$) and $p_0$ are the three momentum of the $K$($K^*$) meson and the energy change of the light quark in B meson rest frame,  respectively:
\begin{eqnarray}
\vert \vec p_{K^{(*)}}\vert &=&{\left [\frac{{(m_B^2+m_{K^{(*)}}^2-m_{\eta'}^2)}^2}{4m_B^2}-m_{K^{(*)}}^2\right ]}^{\frac{1}{2}}\;\; ,\\
p_0&=&\frac{m_B^2-m_b^2}{2m_B}-E_q\;\; ,
\end{eqnarray}
where $E_q$ is the energy of the light quark in $K^{(*)}$ meson.  Inserting $E_q=m_K-m_s^{\rm constituent}\approx 0.05$ GeV in eq. (10) results in
\begin{equation}
BR(B\to K\eta' )=7.0\times 10^{-5}\;\; ,
\end{equation}
and using $E_q\approx m_{K^*}-m_s^{constituent}\approx 0.44 {\rm GeV}\;\;$ in (12) we obtain:\footnote{We estimate $f_{K^*}\approx 205$ GeV$^2$ by using the experimental value of the ratio $\Gamma (\tau\to {K^*}^-\nu_\tau )/\Gamma (\tau\to K^-\nu_\tau )$ and $f_K=0.167$ GeV \cite{pdg}.}:
\begin{equation}
BR(B\to K^*\eta' )=3.4\times 10^{-5}\;\; .
\end{equation}
We note that the results for exclusive decays should not be altered significantly due to momentum dependence of $H$.  This is due to the fact that, unlike the inclusive process, $q^2$ for these decays is fixed at around $1-3$ GeV$^2$.  Measurement of $K^*$ mode will be a crucial testing ground for various mechanisms suggested for $\eta'$ production in hadronic B decays.  For example, our prediction is in contrast to $\Gamma (B\to K^*\eta' )\approx 2\Gamma (B\to K\eta' )$ obtained from the proposed $b\to c\bar cs\to\eta' s$ process\cite{hz}.  We should also emphasize that the ratios of decays $\Gamma (B\to K\eta' )/\Gamma (B\to X_s\eta' )$ and $\Gamma (B\to K^*\eta' )/\Gamma (B\to K\eta' )$, calculated in our mechanism, are independent of the input parameters like $\alpha_s$, $f_B$ and $M_g$, and therefore, free of the uncertainties associated with these parameters. 






\end{document}